\documentclass[12pt]{article}

\usepackage{amssymb}
\usepackage{epsfig}

\begin{document}

\title{
$B \to \rho \gamma$ Decay \\
in the Large Energy Effective Theory
}

\author{{\large Alexander Parkhomenko}\\
        {\it Yaroslavl State University, Yaroslavl, Russia}}

\date{\empty}

\maketitle

\section{Introduction}

There is considerable theoretical interest in the decays $B \to
K^* \gamma$ and $B \to \rho \gamma$ as these processes are 
under intensive experimental  investigations at CLEO, and lately at the 
$B$-factories at KEK and SLAC.

The present measurements of the branching ratios of $B \to K^* \gamma$
decays from the CLEO~\cite{Chen:2001fj}, BABAR~\cite{Tajima:2001qp},
and BELLE~\cite{Aubert:2001me} collaborations yield the following
charge-conjugated world averages~\cite{Hagiwara:pw}:
\begin{eqnarray}
{\cal B}_{\rm exp} (B^\pm \to K^{* \pm} \gamma) =
(3.8 \pm 0.5) \times 10^{-5},
\label{eq:Br-Ks-exp} \\
{\cal B}_{\rm exp} (B^0 \to K^{* 0} \gamma) =
(4.3 \pm 0.4) \times 10^{-5}.
\nonumber
\end{eqnarray}

The Cabibbo-Kobayashi-Maskawa (CKM) disfavored $B \to \rho \gamma$
decays have not yet been observed. The current best limits are from the
BABAR collaboration. One has (at 90\% C.L.)~\cite{Aubert:2002pa}:
\begin{eqnarray}
{\cal B}_{\rm exp} (B^\pm \to \rho^\pm \gamma) < 2.3 \times 10^{-6} ,
\label{eq:BR-exp-BABAR} \\
{\cal B}_{\rm exp} (B^0 \to \rho^0 \gamma) < 1.4 \times 10^{-6} .
\nonumber
\end{eqnarray}
Combined using the isospin symmetry, they yield an improved upper 
limit on the ratio of the branching 
ratios~\cite{Aubert:2002pa}:
\begin{equation}
R_{\rm exp} (\rho \gamma/K^* \gamma) =
\frac{{\cal B}_{\rm exp} (B \to \rho \gamma)}
     {{\cal B}_{\rm exp} (B \to K^* \gamma)}
< 0.05.
\label{eq:ratio-exp}
\end{equation}
Measurement of this ratio will provide  an independent determination of 
the CKM matrix element ratio $\vert V_{td}/V_{ts} \vert$. It has been 
argued in the literature that a  combination of the Heavy Quark Effective 
Theory and the Large Energy Effective Theory (HQET/LEET) provides a sound 
theoretical basis to calculate the branching ratios.

In this paper,  predictions for the branching ratios of the 
$B \to \rho \gamma$ decays are reviewed which have been calculated in the 
HQET/LEET framework. 
The uncertainty connected with the theoretical input (in particular, form 
factors) is considerably reduced in the ratio of the branching ratios:
\begin{equation}
R_{\rm th}(\rho \gamma/K^* \gamma) =
\frac{{\cal B}_{\rm th} (B \to \rho \gamma)}
     {{\cal B}_{\rm th} (B \to K^* \gamma)} .
\label{eq:ratio_th}
\end{equation}
Experimental values of the $B \to K^* \gamma$ branching ratios
in combination with the theoretical estimate of the ratio above
allow to obtain predictions for the $B \to \rho \gamma$ decays 
with reduced uncertainty. The isospin-violating ratio
and the direct CP-asymmetry in the decays $B \to \rho \gamma$  
are also briefly  discussed. More details one can find in 
Ref.~\cite{Ali:2001ez}. 

\section{$B \to \rho \gamma$ decay width in NLO}

The effective Hamiltonian for the $B \to \rho \gamma$ decay (equivalently
$b \to d \gamma$ process) at the scale $\mu = O (m_b)$, where~$m_b$ is the
$b$-quark mass, is as follows:
\begin{eqnarray}
{\cal H}_{\rm eff} & = & \frac{G_F}{\sqrt 2}
\left \{
V_{ub} V_{ud}^* \,
\left [
C_1 (\mu) \, {\cal O}_1^{(u)} (\mu) +
C_2 (\mu) \, {\cal O}_2^{(u)} (\mu)
\right ]
\right.
\nonumber \\
& + &
V_{cb} V_{cd}^* \,
\left [
C_1 (\mu) \, {\cal O}_1^{(c)} (\mu) +
C_2 (\mu) \, {\cal O}_2^{(c)} (\mu)
\right ]
\label{eq:Hamiltonian} \\
& - &
\left.
V_{tb} V_{td}^* \,
\left [
C_7^{\rm eff} (\mu) \, {\cal O}_7 (\mu) +
C_8^{\rm eff} (\mu) \, {\cal O}_8 (\mu)
\right ]
+ \ldots
\right \} ,
\nonumber
\end{eqnarray}
where the set of operators is ($q = u, c$):
\begin{eqnarray}
{\cal O}_1^{(q)} & = &
(\bar d_\alpha \gamma_\mu (1 - \gamma_5) q_\beta) \,
(\bar q_\beta \gamma^\mu (1 - \gamma_5) b_\alpha) ,
\label{eq:operator-O1} \\
{\cal O}_2^{(q)} & = &
(\bar d_\alpha \gamma_\mu (1 - \gamma_5) q_\alpha) \,
(\bar q_\beta \gamma^\mu (1 - \gamma_5) b_\beta) ,
\label{eq:operator-O2} \\
{\cal O}_7 & = & \frac{e m_b}{8 \pi^2} \,
(\bar d_\alpha \sigma^{\mu \nu} (1 + \gamma_5) b_\alpha) \,
F_{\mu \nu} ,
\label{eq:operator-O7} \\
{\cal O}_8 & = & \frac{g_s m_b}{8 \pi^2} \,
(\bar d_\alpha \sigma^{\mu \nu} (1 + \gamma_5)
T^A_{\alpha \beta} b_\beta) \, G^A_{\mu \nu} .
\label{eq:operator-O8}
\end{eqnarray}
The strong and electroweak four-quark penguin operators are
assumed to be present in the effective Hamiltonian~(\ref{eq:Hamiltonian}) 
and denoted by ellipses there. They are not taken into account 
due to the small values of the corresponding Wilson coefficients. 

The effective Hamiltonian~(\ref{eq:Hamiltonian}) sandwiched between
the~$B$- and $\rho$-meson states can be expressed in terms of
matrix elements of bilinear quark currents defining a heavy-light
transition. These matrix elements are dominated by strong
interactions at small momentum transfer and cannot be calculated
perturbatively. The general decomposition of the matrix elements on all
possible Lorentz structures admits seven scalar functions (form factors):
$V^{(\rho)}$, $A_i^{(\rho)}$ ($i = 0, 1, 2$), and~$T_i^{(\rho)}$
($i = 1, 2, 3$) of the momentum squared~$q^2 = (p_B - p_\rho)^2$
transferred from the heavy meson to the light one~\cite{Beneke:2000wa}:
\begin{eqnarray}
&&
\left < \rho (p_\rho, \varepsilon^*) |
\bar d \, \gamma^\mu b
| \bar B (p_B) \right > =
\frac{2 i \, V^{(\rho)} (q^2)}{m_B + m_\rho} \,
{\rm eps} (\mu, \varepsilon^*, p_\rho, p_B) ,
\label{eq:vertor-current} \\
&&
\left < \rho (p_\rho, \varepsilon^*) |
\bar d \, \gamma^\mu \gamma_5 q_\nu b
| \bar B (p_B) \right > =
2 m_\rho \, A_0^{(\rho)} (q^2) \,
\frac{(\varepsilon^* q)}{q^2} \, q^\mu
\label{eq:axial-current} \\
&&
\hspace{25mm}
+ A_1^{(\rho)} (q^2) \, (m_B + m_\rho)
\left [ \varepsilon^{* \mu} -
\frac{(\varepsilon^* q)}{q^2} \, q^\mu \right ]
\nonumber \\
&&
\hspace{25mm}
- A_2^{(\rho)} (q^2) \, \frac{(\varepsilon^* q)}{m_B + m_\rho} \,
\left [ (p_B + p_\rho)^\mu - \frac{(m_B^2 - m_\rho^2)}{q^2} q^\mu
\right ] ,
\nonumber \\
&&
\left < \rho (p_\rho, \varepsilon^*) |
\bar d \, \sigma^{\mu \nu} q_\nu b
| \bar B (p_B) \right > =
2 \, T_1^{(\rho)} (q^2) \, {\rm eps} (\mu, \varepsilon^*, p_\rho, p_B) ,
\label{eq:tensor-current} \\
&&
\left < \rho (p_\rho, \varepsilon^*) |
\bar d \, \sigma^{\mu \nu} \gamma_5 q_\nu b
| \bar B (p_B) \right > = 
\label{eq:axial-tensor-current} \\
&&
\hspace{25mm}
- i \, T_2^{(\rho)} (q^2) \, [(m_B^2 - m_\rho^2) \,
\varepsilon^{* \mu} - (\varepsilon^* q) \, (p_B + p_\rho)^\mu]
\nonumber \\
&&
\hspace{25mm}
- i \, T_3^{(\rho)} (q^2) \, (\varepsilon^* q) \,
\left [
q^\mu - \frac{q^2}{m_B^2 - m_\rho^2} \, (p_B + p_\rho)^\mu
\right ] ,
\nonumber
\end{eqnarray}
where ${\rm eps} (\mu, \varepsilon^*, p_\rho, p_B) =
\varepsilon^{\mu \nu \alpha \beta} \varepsilon^*_\nu p_{\rho
\alpha} p_{B \beta}$. The heavy quark symmetry and the behavior
of the final meson in the large energy limit (the large recoil
limit) allow to reduce the number of independent form factors to
two only: $\xi^{(\rho)}_\perp (q^2)$ and $\xi^{(\rho)}_\| (q^2)$.
The $B \to \rho \gamma$ decay amplitude is proportional to one of
them~-- $\xi^{(\rho)}_\perp (q^2)$, which is related to the form
factors introduced above for the case $q^2 = 0$ as follows (terms
of order $m_\rho^2/m_B^2$ are neglected):
\begin{equation}
\frac{m_B}{m_B + m_\rho} \, V^{(\rho)} (0) =
\frac{m_B + m_\rho}{m_B} \, A_1^{(\rho)} (0) =
T_1^{(\rho)} (0) = T_2^{(\rho)} (0) = \xi^{(\rho)}_\perp (0) .
\label{eq:FF-relation}
\end{equation}
These relations among the form factors in the symmetry
limit are broken by perturbative QCD radiative corrections
arising from the vertex renormalization and the hard spectator
interactions.
To incorporate both types of QCD corrections, a tentative
factorization formula for the heavy-light form factors at large
recoil and at leading order in the inverse heavy meson mass was
introduced~\cite{Beneke:2000wa}:
\begin{equation}
F^{(\rho)}_k  = C_{\perp k} \xi^{(\rho)}_\perp + 
\Phi_B \otimes T_k \otimes \Phi_\rho ,
\label{eq:factor-general}
\end{equation}
where $F^{(\rho)}_k$ is any of the four form factors in the 
$B \to \rho$ transitions related by Eq.~(\ref{eq:FF-relation}), 
$C_{\perp k} = C_{\perp k}^{(0)} [1 + O (\alpha_s)]$ are the
renormalization coefficients, $T_k$ is a hard-scattering kernel
calculated in $O (\alpha_s)$,
$\Phi_B$ and~$\Phi_\rho$ are the light-cone distribution
amplitudes of the~$B$- and $\rho$-mesons convoluted with 
the kernel~$T_k$.

In the leading order the electromagnetic penguin operator ${\cal O}_7$
contributes in the $B \to \rho \gamma$ decay amplitude at the tree level.
Taking into account the definitions of the $B \to \rho$ transition
form factors in the tensor~(\ref{eq:tensor-current}) and the 
axial-tensor~(\ref{eq:axial-tensor-current}) currents and the 
symmetry relation $T^{(\rho)}_1 (0) = T^{(\rho)}_2 (0)$, 
the amplitude has the form:
\begin{eqnarray}
M^{(0)} & = & - \frac{G_F}{\sqrt 2} \, V_{tb} V_{td}^* \,
\frac{e \bar m_b (\mu)}{4 \pi^2} \, C_7^{(0) {\rm eff}} (\mu) \,
T^{(\rho)}_1 (0)
\label{eq:ME0} \\
& \times &
\left [ (P q) (e^* \varepsilon^*) - (e^* P) (\varepsilon^* q)
+ i \, {\rm eps} (e^*, \varepsilon^*, P, q)
\right ] ,
\nonumber
\end{eqnarray}
where $q = p_B - p_\rho$ and $e^*$ are the photon four-momentum
and polarization vector, respectively, and $P = p_B + p_\rho$.

The branching ratio can be easily obtained and results in the form:
\begin{equation}
{\cal B}_{\rm th}^{\rm LO} (B \to \rho \gamma) = \tau_B \,
\frac{G_F^2 \alpha |V_{tb} V_{td}^*|^2 m_B^3}{32 \pi^4} \,
\left [ 1 - \frac{m_\rho^2}{m_B^2} \right ]^3
\bar m_b^2 (\mu) \, |C_7^{(0) {\rm eff}} (\mu)|^2 \, 
|T^{(\rho)}_1 (0)|^2 .
\label{eq:Br-rho}
\end{equation}
It is natural to assume the $\mu$-dependence of the form factor,
$T^{(\rho)}_1 (0, \mu)$, for compensating the dependence on the 
scale~$\mu$ originated by the $b$-quark mass, $\bar m_b (\mu)$, 
and the Wilson coefficient, $C_7^{(0) {\rm eff}} (\mu)$, in the 
branching ratio.

The branching ratio of the $B \to K^* \gamma$ decays can be easily
obtained from Eq.~(\ref{eq:Br-rho}) by replacing: $V_{td} \to V_{ts}$,
$m_\rho \to m_{K^*}$, and $T^{(\rho)}_1 (0) \to T^{(K^*)}_1 (0)$, which
yield the ratio of branching ratios~(\ref{eq:ratio_th}) as:
\begin{equation}
R_{\rm th}^{(0)} (\rho \gamma/K^* \gamma) =
S_\rho \left | \frac{V_{td}}{V_{ts}} \right |^2
\left [ \frac{m_B^2 - m_\rho^2}{m_B^2 - m_{K^*}^2} \right ]^3
\left | \frac{T^{(\rho)}_1 (0, \mu)}{T^{(K^*)}_1 (0, \mu)} \right |^2 ,
\label{eq:ratio0_result}
\end{equation}
where $S_\rho = 1$ for the charged $\rho$-meson and $S_\rho = 1/2$ for
the neutral one.

There is also contribution from the annihilation diagrams to the
$B \to \rho \gamma$ decay width. This additional contribution
modifies the equation above as follows:
\begin{equation}
R_{\rm th} (\rho \gamma/K^* \gamma) =
R_{\rm th}^{(0)} (\rho \gamma/K^* \gamma)
\left [ 1 + \Delta R (\rho/K^*) \right ] .
\label{eq:ratio_ann_result}
\end{equation}
The radiation from quarks inside such a meson is compensated by the
diagram with the photon emitted from the vertex (for a recent review
of this topic see Ref.~\cite{Khodjamirian:2001ga}). Only one
annihilation diagram with the photon emitted from the spectator
quark in the $B$-meson is numerically important  and its strength
can be parameterized by the dimensionless factor~$\varepsilon_A$:
\begin{eqnarray}
&&
\Delta R (\rho/K^*) = \lambda_u \, \varepsilon_A ,
\label{eq:ann_contribution} \\
&&
\lambda_u = \frac{V_{ub} V_{ud}^*}{V_{tb} V_{td}^*} =
- \left | \frac{V_{ub} V_{ud}^*}{V_{tb} V_{td}^*} \right |
{\rm e}^{i \alpha}
= F_1 + i F_2 ,
\label{eq:CKM-ratio}
\end{eqnarray}
where~$\alpha$ is one of the angle from the unitarity triangle.
In the neutral $B$-meson decays the parameter~$\varepsilon_A$ is 
numerically small and can be neglected at the accuracy accepted.
For the charged $B$-meson decays the LCSR value   
$\varepsilon_A = 0.3 \pm 0.1$~\cite{Ali:vd} is used in the analysis. 

There is also QCD corrections [of order $O (\alpha_s)$] which 
are called further as the next-to-leading order (NLO) ones. 
The total NLO correction to the $B \to \rho \gamma$ decay width 
consists of: 

$\bullet$ \underline{$b$-quark mass $\bar m_b (\mu)$.}
In the modified minimal subtraction scheme at the renormalization
scale~$\mu$ it can be connected with the $b$-quark pole mass,
$m_{b, {\rm pole}}$, by the relation:
\begin{equation}
\bar m_b (\mu) = m_{b, {\rm pole}}
\left [ 1 + \frac{\alpha_s (\mu)}{4 \pi} \, C_F
\left ( 3 \ln \frac{m_{b, {\rm pole}}^2}{\mu^2}  - 4 \right ) \right ].
\end{equation}

$\bullet$ \underline{Wilson coefficient $C_7^{\rm eff} (\mu)$.}
\begin{equation}
C_7^{\rm eff} (\mu) = C_7^{(0) {\rm eff}} (\mu) +
\frac{\alpha_s (\mu)}{4 \pi} \, C_7^{(1) {\rm eff}} (\mu).
\end{equation}
The explicit expressions for the Wilson coefficient can be found
in Ref.~\cite{Chetyrkin:1996vx}.

$\bullet$
\underline{The factorizable NLO corrections to the form factors.}
These corrections are described by the diagrams with the 
${\cal O}_7$-operator~(\ref{eq:operator-O7}). They can be divided 
into the vertex and hard-spectator corrections~\cite{Beneke:2000wa}:
\begin{equation}
T^{(\rho)}_1 (0, \mu) = \xi^{(\rho)}_\perp (0) \left [ 1 +
\frac{\alpha_s (\mu)}{4 \pi} \,
C_F \left ( \ln \frac{m_{b, {\rm pole}}^2}{\mu^2} - 1 \right)
+ \frac{\alpha_s (\mu_{\rm sp})}{4 \pi} \, C_F \, 
\frac{\Delta F_\perp (\mu_{\rm sp})}{2 \xi^{(\rho)}_\perp (0)}
\right ] ,
\label{eq:T1-NLO-fact}
\end{equation}
where $\xi^{(\rho)}_\perp (0)$ is the value of the $T^{(\rho)}_1 (0)$
form factor in the HQET/LEET limit,
$\mu_{\rm sp} = \sqrt{\mu \Lambda_{\rm H}}$ ($\Lambda_{\rm H} \simeq
0.5$~GeV) is the typical scale of the gluon virtuality in the
hard-spectator corrections, and $\Delta F_\perp^{(\rho)}$ is the 
dimensionless quantity which defines the strength of the 
hard-spectator corrections:
\begin{equation}
\Delta F^{(\rho)}_\perp (\mu_{\rm sp}) =
\frac{8 \pi^2 f_B f_\perp^{(\rho)} (\mu_{\rm sp})}{3 m_B \lambda_{B,+}}
<\bar u^{-1}>_\perp^{(\rho)} (\mu_{\rm sp})
\simeq 1.64 .
\end{equation}
The estimation was done on the scale of the spectator
corrections $\mu_{\rm sp} = 1.52$~GeV~\cite{Ali:2001ez}.

$\bullet$ \underline{The nonfactorizable NLO corrections.}
They are also of two types: the vertex and the hard-spectator corrections.
The nonfactorizable vertex corrections can be taken from inclusive
$B \to X_d \gamma$ decay~\cite{Ali:1998rr}.
The nonfactorizable hard-spectator ones were recently calculated by 
several groups~\cite{Ali:2001ez,Beneke:2001at,Bosch:2001gv}. 

The total contribution to the form factor originated by the 
hard-spectator corrections is~\cite{Ali:2001ez}:
\begin{eqnarray}
\Delta_{\rm sp} T_1^{(\rho)} (0, \mu)
& \simeq &
\frac{\alpha_s (\mu)}{4 \pi} \, C_F \,
\frac{\Delta F_\perp^{(\rho)} (\mu)}{2}
\left [
1 + \frac{C_8^{(0){\rm eff}} (\mu)}{3 C_7^{(0){\rm eff}} (\mu)}
\right.
\label{eq:DT1-spectator} \\
& + &
\left. \frac{C_2^{(0)} (\mu)}{3 C_7^{(0){\rm eff}} (\mu)}
\left (
1 + \frac{V_{cd}^* V_{cb}}{V_{td}^* V_{tb}} \,
\frac{h^{(\rho)} (z, \mu)}
     {\left < \bar u^{-1} \right >_\perp^{(\rho)} (\mu)}
\right )
\right ] ,
\nonumber
\end{eqnarray}
where $h^{(\rho)} (z, \mu)$ is the complex function of the quark
mass ratio $z = m_c^2/m_b^2$ originated by the $c$-quark loop
which analytic expression can be found in Ref.~\cite{Ali:2001ez}.

The NLO corrections discussed above  modify the $B \to \rho \gamma$
branching ratio and the result for the charged-conjugate averaged
branching ratio can be written in the form:
\begin{eqnarray}
&& \hspace*{-10mm}
\bar {\cal B}_{\rm th} (B^\pm \to \rho^\pm \gamma) =
\tau_{B^+} \, \frac{G_F^2 \alpha |V_{tb} V_{td}^*|^2}{32 \pi^4} \,
m_{b,{\rm pole}}^2 \, m_B^3
\left [ 1 - \frac{m_\rho^2}{m_B^2} \right ]^3
\left [ \xi_\perp^{(\rho)} (0) \right ]^2 \,
\label{eq:DecayWidth} \\
&& \hspace*{-10mm}
\times
\left \{ (C^{(0){\rm eff}}_7 + A^{(1)t}_R)^2 +
(F_1^2 + F_2^2) \, (A^u_R + L^u_R)^2
+ 2 F_1 \, [ C^{(0){\rm eff}}_7 (A^u_R + L^u_R)
        + A^{(1)t}_R L^u_R ]
\right \} ,
\nonumber
\end{eqnarray}
where $L^u_R = \epsilon_A \, C^{(0) {\rm eff}}_7$.
The NLO amplitude~$A^{(1)t} (\mu)$ of the decay presented here can 
be decomposed in three contributing parts~\cite{Ali:2001ez}:
\begin{equation}
A^{(1)t} (\mu) = A^{(1)}_{C_7} (\mu) +
A^{(1)}_{\rm ver} (\mu) + A^{(1)\rho}_{\rm sp} (\mu_{\rm sp}), 
\label{eq:NLO-ampl-decomposition}
\end{equation}
where the correction due to the $b$-quark mass is included in the 
$A^{(1)}_{\rm ver} (\mu)$ part. 
The amplitude~$A^{(1) K^*} (\mu)$ for the $B \to K^* \gamma$ decay
can be written in a similar form and differs from~$A^{(1)t}$ by the 
hard-spectator part~$A^{(1) K^*}_{\rm sp} (\mu)$ only~\cite{Ali:2001ez}. 
Note that the $u$-quark contribution~$A^u (\mu)$ from the penguin diagrams, 
which also involves the contribution of hard-spectator corrections, 
can not be ignored in the $B \to \rho \gamma$ decay.

Using the presentation~(\ref{eq:DecayWidth}) of the branching ratio,
the dynamical function $\Delta R (\rho/K^*)$, defined by
Eq.~(\ref{eq:ratio_ann_result}), in the NLO and with the annihilation
contribution taken into account can be written:
\begin{eqnarray}
&& \Delta R (\rho/K^*) =
\left [ 2 \epsilon_A \, F_1 + \epsilon_A^2 (F_1^2 + F_2^2) \right ] 
\left ( 1 - \frac{2 A^{(1)K^*}}{C^{(0) {\rm eff}}_7} \right ) - 
\frac{2 A^{(1)K^*}}{C^{(0) {\rm eff}}_7}  
\label{eq:DR-NLO} \\
&&  \hspace*{10mm}
+ \frac{2}{C^{(0) {\rm eff}}_7} \, {\rm Re}
\left [
 A^{(1)\rho}_{\rm sp} -  A^{(1)K^*}_{\rm sp}
+ F_1 (A^u + \epsilon_A A^{(1)t}) + \epsilon_A (F_1^2 + F_2^2) A^u
\right ] .
\nonumber
\end{eqnarray}

\section{Phenomenology of $B \to \rho \gamma$ Decays}

\paragraph{$B \to \rho \gamma$ Branching Ratios.}
For numerical predictions of the $B \to \rho \gamma$ branching ratios
it is better to use ratio of the $B \to \rho \gamma$ and
$B \to K^* \gamma$ decay widths~(\ref{eq:ratio_th}) and then connect it
with the experimentally measured values of $B \to K^* \gamma$ branching
ratios~(\ref{eq:Br-Ks-exp}).

To do this, let us start with the discussion of form factors.
$SU_F (3)$-breaking effects in the QCD form factors $T_1^{(K^*)} (0)$
and $T_1^{(\rho)} (0)$ have been evaluated within the QCD
sum-rules~\cite{Ali:vd}. These can be taken to hold also for the ratio
of the HQET form factors. Thus, we take $\zeta = \xi_\perp^{(\rho)} (0) /
\xi_\perp^{(K^*)} (0) \simeq 0.76 \pm 0.06$.
As it was pointed out in Ref.~\cite{Ali:2002kw}, the error here
is not on~$\zeta$ by itself, but rather on the deviation of~$\zeta$
from its $SU_F (3)$-symmetry limit, i.~e. $1 - \zeta$.

The main uncertainties in the dynamical functions~$\Delta R(\rho/K^*)$
come from the uncertainties in the CKM angle~$\alpha$ and the
nonperturbative pa\-ra\-me\-ters~$\xi^{(\rho)}_\perp (0)$
and~$\xi^{(K^*)}_\perp (0)$. Taking into account various parametric
uncertainties, it is found that the dynamical functions $\Delta R(\rho/K^*)$
are constrained in the range~\cite{Ali:2001ez}:
\begin{equation}
\left | \Delta R(\rho^\pm/K^{*\pm}) \right | \le 0.25,
\qquad
\left | \Delta R(\rho^0/K^{*0}) \right | \le 0.13,
\end{equation}
with the central values $\Delta R (\rho^\pm/K^{*\pm}) \simeq 
\Delta R (\rho^0/K^{*0}) \simeq 0$. This quantifies the statement 
that the ratio $R_{\rm th} (\rho \gamma/K^* \gamma)$ is stable 
against~$O(\alpha_s)$ and $1/m_B$-corrections.

Taking into account the ratio of the CKM matrix elements:
$ \left | V_{td}/V_{ts} \right | = 0.194 \pm 0.029$,
the branching ratios can be estimated as~\cite{Ali:2001ez}
\begin{eqnarray}
&& \bar {\cal B}_{\rm th} (B^\pm \to \rho^\pm \gamma) =
(0.90 \pm 0.33 [{\rm th}] \pm 0.10 [{\rm exp}]) \times 10^{-6},
\nonumber \\
&& \bar {\cal B}_{\rm th} (B^0 \to \rho^0 \gamma) =
(0.49 \pm 0.18 [{\rm th}] \pm 0.04 [{\rm exp}]) \times 10^{-6},
\nonumber
\end{eqnarray}
where the SM favoured range $77^\circ \le \alpha \le 113^\circ$
was used.
In the above estimates, the first error is due to the uncertainties
of the theory and the second is from the  experimental data
on the $B \to K^* \gamma$ branching ratios. 
The recent experimental upper limits on these decays 
by the BABAR collaboration~(\ref{eq:BR-exp-BABAR}) are
approximately a factor three above the predicted ones. We expect that
the BABAR and BELLE experiments will soon reach the SM sensitivity in 
these decays.

\begin{figure}[tb]
\centerline{\epsfysize=.35\textheight
            \epsffile{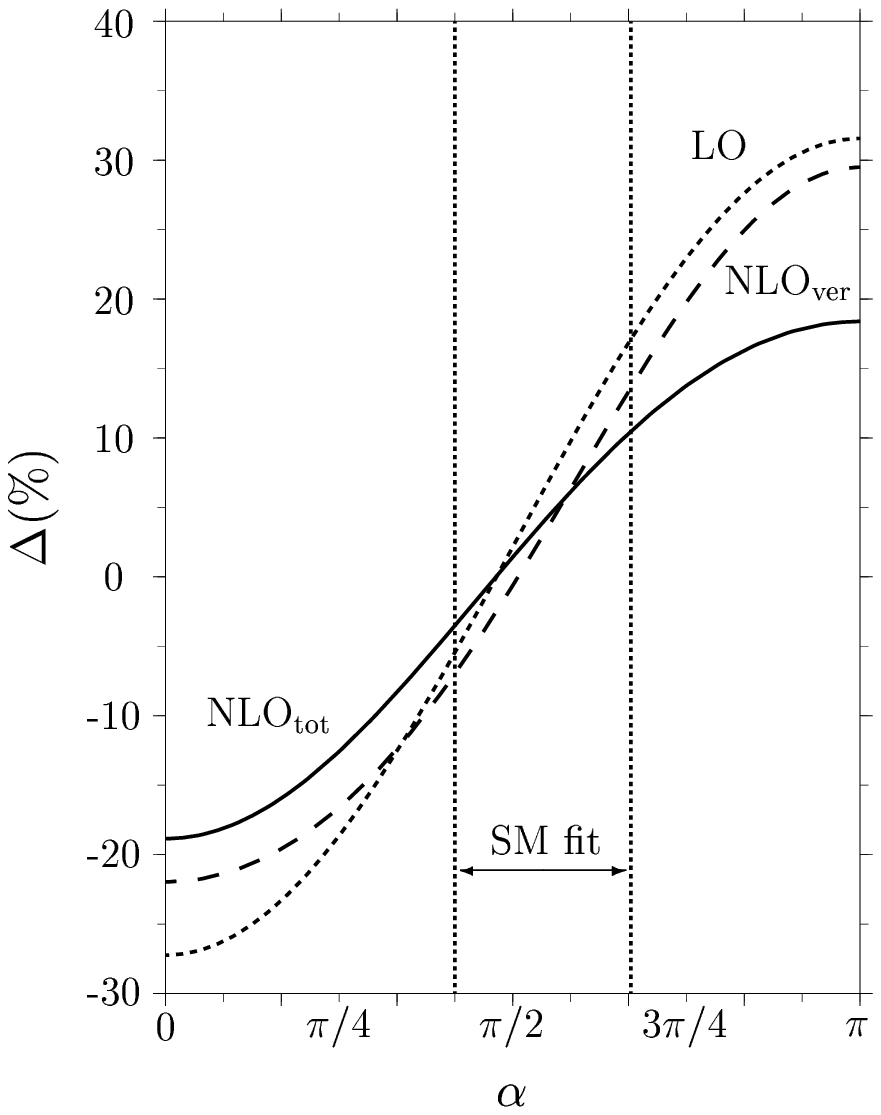}}
\caption{The charge-conjugate averaged ratio~$\Delta$ for $B \to \rho
         \gamma$ decays as a function of the unitarity triangle
          angle~$\alpha$ in the leading order (dotted curve),
          next-to-leading order without (dashed curve) and
         with (solid curve) hard-spectator corrections. The $\pm 1 \sigma$
         allowed band of~$\alpha$ from the SM unitarity fits is also
         indicated.}
\label{fig:Delta}
\end{figure}

\paragraph{Isospin-Violating Ratios.}
The numerical analysis is presented for the charge-conjugate averaged
of the isospin-violating ratios in the $B \to \rho \gamma$ decays:
\begin{equation}
\Delta = \frac{1}{4} \,
\left [
\frac{\Gamma (B^- \to \rho^- \gamma)}
     {\Gamma (\bar B^0 \to \rho^0 \gamma)}
+ \frac{\Gamma (B^+ \to \rho^+ \gamma)}
       {\Gamma (B^0 \to \rho^0 \gamma)}
\right ] - 1 .
\label{eq:Delta}
\end{equation}
The dependence on the unitarity triangle angle~$\alpha$ is presented
in Fig.~\ref{fig:Delta}.
%
%
%
%
The charge-conjugate average~$\Delta$ for the $B \to \rho \gamma$ 
decays is found to be likewise stable against the NLO and
$1/m_B$-corrections~\cite{Ali:2001ez}.
In the expected range of the CKM parameters, this quantity 
is inside the interval $|\Delta| \leq 10\%$.

\begin{figure}[tb]
\centerline{\epsfxsize=.35\textwidth
            \epsffile{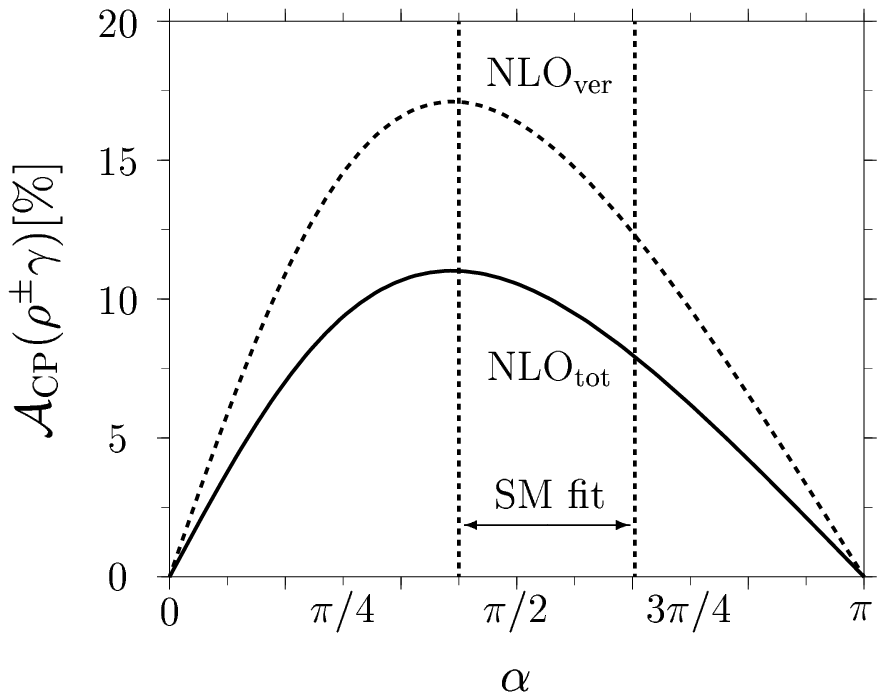} \qquad
            \epsfxsize=.35\textwidth
            \epsffile{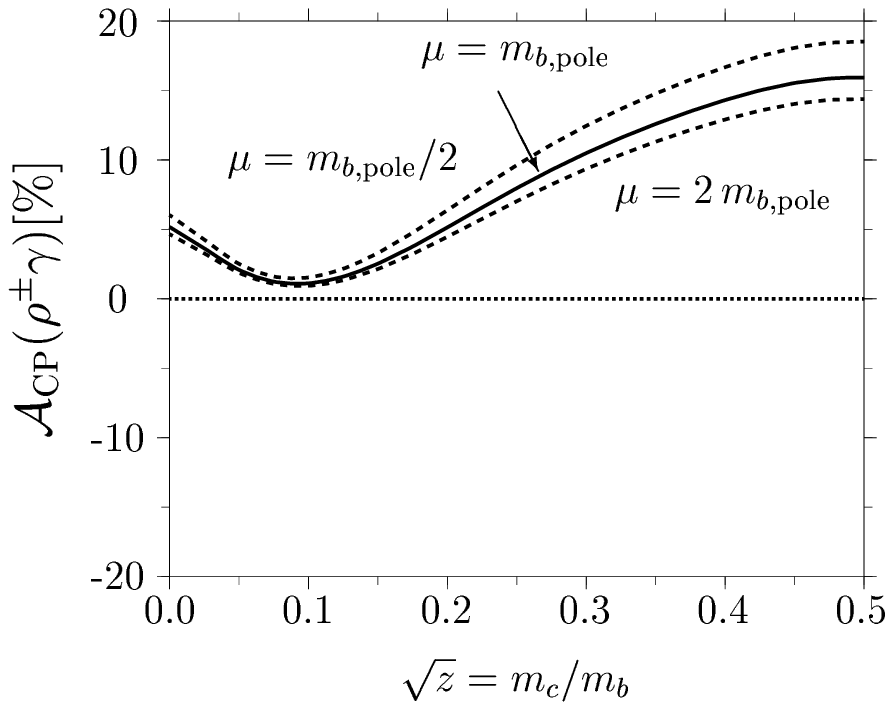}}
\caption{Left figure: Direct CP-asymmetry in the decays
         $B^\pm \to \rho^\pm \gamma$ as a function of the unitarity
         triangle angle~$\alpha$ without (dotted curves) and with
         (solid curves) the hard-spectator corrections.
         The $\pm 1 \sigma$ allowed
         band of $\alpha$ from the SM unitarity fits is also indicated.
         Right figure: Direct CP-asymmetry in the decays
         $B^\pm \to \rho^\pm \gamma$ as a function of the quark mass
         ratio~$\sqrt z = m_c/m_b$; the scale dependence of the
         asymmetry is shown in the interval:
         $m_{b,{\rm pole}}/2 \le \mu \le 2 m_{b,{\rm pole}}$.}
\label{fig:ACPdir}
\end{figure}

\paragraph{Direct CP-Asymmetry.}
The direct CP-asymmetry in the $B^\pm \to \rho^\pm \gamma$ decay rates
is defined as follows:
\begin{equation}
{\cal A}_{\rm CP} (\rho^\pm \gamma) =
\frac{{\cal B} (B^- \to \rho^- \gamma) - {\cal B} (B^+ \to \rho^+ \gamma)}
  {{\cal B} (B^- \to \rho^- \gamma) + {\cal B} (B^+ \to \rho^+ \gamma)} .
\label{eq:CP-asymmetry}
\end{equation}
The CP-asymmetry ${\cal A}_{\rm CP} (\rho^\pm \gamma)$ receives
contributions from the hard-spectator corrections which tend to
decrease its value estimated from the vertex corrections alone.
The dependencies of the CP-asymmetry on the angle~$\alpha$ and on
the quark mass ratio~$\sqrt z = m_c/m_b$ are presented in
Fig.~\ref{fig:ACPdir}.
%
%
%
%
The Standard Model estimates show that the direct CP-asymmetry is
definitely positive and for $0.2 \lesssim \sqrt z \lesssim 0.3$ 
is inside the interval:
$5\% < {\cal A}_{\rm CP} (\rho^\pm \gamma) < 15\%$.

\medskip

\paragraph{Acknowledgements.} It is a great pleasure to thank 
Ahmed Ali for the fruitful collaboration and useful remarks on 
the manuscript.

\end{document}